# Magnetism in $J_{eff} = 1/2$ Kagome Antiferromagnet: Thermodynamics, Nuclear Magnetic Resonance, Muon Spin Resonance, and Inelastic Neutron Scattering Studies

A. Yadav[1], A. Elghandour[2], T. Arh[3], D. T. Adroja[4,5], M. D. Le[5], G. B. G. Stenning[5], M. Aouane[5], S. Luther[6], F. Hotz[7], T. J. Hicken[7], H. Luetkens[7], A. Zorko[3,8], R. Klingeler[2], and P. Khuntia[1,9, *]

[1]*Department of Physics, Indian Institute of Technology Madras, Chennai 600036, India*
[2]*Kirchhoff Institute of Physics, Heidelberg University, INF 227, D-69120 Heidelberg, Germany*
[3]*Jožef Stefan Institute, Jamova c. 39, SI-1000 Ljubljana, Slovenia*
[4]*ISIS Facility, Rutherford Appleton Laboratory, Chilton, Didcot, Oxon, OX11 0QX, United Kingdom*
[5]*Highly Correlated Matter Research Group, Physics Department, University of Johannesburg, Auckland Park 2006, South Africa*
[6]*Dresden High Magnetic Field Laboratory (HLD-EMFL), Helmholtz-Zentrum Dresden-Rossendorf, 01328 Dresden, Germany*
[7]*Laboratory for Muon Spin Spectroscopy, Paul Scherrer Institute, CH-5232 Villigen, Switzerland*
[8]*Faculty of Mathematics and Physics, University of Ljubljana, Jadranska u. 19, SI-1000 Ljubljana, Slovenia*
[9]*Quantum Centre of Excellence for Diamond and Emergent Materials, Indian Institute of Technology Madras, Chennai 600036, India.*

(Dated: 18 December, 2024)

The intertwining between competing degrees of freedom, anisotropy, and frustration-induced strong quantum fluctuations offers an ideal ground to realize exotic quantum phenomena in the rare-earth-based kagome lattice. Herein, we report synthesis, structure, thermodynamic, muon spin relaxation (μSR), nuclear magnetic resonance (NMR), and inelastic neutron scattering (INS) studies on a frustrated quantum magnet $Nd_3BWO_9$ wherein $Nd^{3+}$ ions constitute a distorted kagome lattice. The INS experiments on $Nd_3BWO_9$ allow us to establish a detailed crystal electric field spectrum. The magnetic susceptibility reveals the presence of two energy scales in agreement with the INS results wherein the higher energy state is dominated by thermal population of crystal-electric-field excitations. The lowest Kramers ground-state doublet is well separated from the excited state suggesting that the compound realizes a low-energy $J_{eff}$ =1/2 state at low temperatures. The low energy state is witnessed via thermodynamic results that reveal an anomaly at 0.3 K typical of a phase transition, which is attributed to the presence of complex magnetic ordering phenomena and the broad maximum in the specific heat well above 0.3 K indicates the presence of short-range spin correlations. The isothermal magnetization reveals a field-induced 1/3 magnetization plateau at low temperatures. μSR relaxation rate experiments, on the other hand, neither show the signature of a phase transition nor spin-freezing down to 34 mK. The zero field μSR relaxation rate is governed by an Orbach process and reveals the presence of a fluctuating state owing to depopulation of crystal field levels reflected as a constant value of relaxation rate in the temperature range $0.04 ≤ T ≤ 10$ K. NMR results indicate the presence of fluctuating $Nd^{3+}$ moments down to 1.8 K consistent with μSR experiments. Our comprehensive results reveal that a field-induced quantum phenomenon is at play, exemplifying the proximity effect of competing magnetic states and the coexistence of static and fluctuating moments , along with short range spin correlations in this frustrated kagome magnet. The broad rare-earth $RE_3BWO_9$ family of frustrated kagome magnets is a promising candidate to host exotic quantum states driven by spin-orbit coupling and frustration.

---

*pkhuntia@iitm.ac.in

# I. INTRODUCTION

Geometrically frustrated magnets, wherein the incompatibility of exchange interactions between the spins in minimizing the ground state energy lead to macroscopic ground state degeneracy, non-trivial low-energy excitation spectra, and quantum phase transitions, that represent a mainstream field of modern condensed matter [1]. Quantum mechanical effects such as spin-orbit coupling, entanglement, and quantum fluctuations, as well as the non-trivial topology of the electron wave function of frustrated quantum materials, offer an incredible ground to realize intriguing quantum and topological states such as quantum spin liquid (QSL) with exotic quasi-particle excitations [2, 3]. QSL is a highly entangled state of quantum matter wherein frustration-induced strong quantum fluctuations evade magnetic long-range ordering down to absolute zero temperature despite strong exchange interaction between spins. QSL is promising to host fractional excitations that are coupled to emergent gauge fields. QSL has the potential to address some of the recurring themes in quantum condensed matter and is highly relevant for topological quantum computing [1–3]. In this context, the frustrated kagome antiferromagnet with corner-sharing triangles of magnetic moments renders an emblematic two-dimensional (2D) model that can maximize frustration-induced quantum fluctuations and is a promising contender to host exotic quantum phenomena [5–10]. In particular, quantum spin liquids [11-13], magnetization plateaus [14], incommensurate magnetic order [15], and chiral spin-ordered states [16], massive Dirac fermions in ferromagnetic kagome metal, anomalous Hall effect [17–19], Berry phase [20], and unconventional superconductivity [21–22] in kagome magnets represent an appealing track in contemporary quantum matter to establish theoretical models that could resemble the properties of experimental candidates [23, 24]. A weak perturbation in the nearest-neighbor Heisenberg model, such as next-nearest neighbor interactions, anisotropy, including the Dzyaloshinskii-Moriya interaction, in kagome magnets, can have a striking effect on the underlying spin Hamiltonian and ground state properties. [1,4, 25–34. Quantum criticality is also at play in diverse QSL models and frustrated magnets, which suggests that non-thermal external tuning parameters, i.e., pressure and magnetic field, can drive a quantum phase transition at $T\rightarrow 0$, exemplifying the competition between the nearly degenerate phases that have very similar characteristic energy scales [4–6,35].

The rare-earth-based frustrated kagome quantum magnet with distinct magnetic ions is a promising candidate to stabilize a structurally perfect lattice comprising a highly tunable nature of weak exchange interactions mediated by strongly localized magnetic moments. In these kagome materials, the synergistic interplay between strong spin-orbit coupling and crystal electric field (CEF) often generates an exchange anisotropy, which often leads to effective spin-1/2 degrees of freedom that are prone to strong quantum fluctuations at low temperatures, thus offering an excellent route to stabilize many-body quantum states [2–7, 36–43]. However, defects and anti-site disorder in many experimental frustrated magnets pose a strong constraint for the unambiguous identification of the novel quantum state and associated magnetic excitations. The current challenge is to design, discover, and investigate structurally perfect kagome antiferromagnet representatives that provide a viable platform to realize fascinating quantum states. In this vein, the recently discovered rare-earth family $RE_3BWO_9$ (RE = rare-earth elements) where $RE^{3+}$ ions form a distorted-kagome lattice in the *ab*-plane and are stacked along the z-axis provides an alternate route to realize spin-orbit driven correlated quantum states with non-trivial low-energy excitation spectra [44]. For instance, $RE_3BWO_9$ (RE = Gd, Dy, and Ho) shows a large magnetocaloric effect under the application of an external magnetic field, which is highly relevant for magnetic refrigeration at low temperatures. $Pr_3BWO_9$ exhibits a dynamic ground state down to 0.03 K with Ising-like spin correlations, while $Sm_3BWO_9$ hosts an incommensurate magnetic order. $Nd_3BWO_9$ is also a promising candidate to host frustration and field-induced quantum states, including a 1/3 magnetization plateau, and unconventional quantum critical phenomena at low temperatures [45-50]. However, microscopic details concerning spin-freezing, local magnetic field distribution at low temperatures, and a comprehensive crystal electric field scheme are not yet clear for the kagome magnet $Nd_3BWO_9$. Moreover, detailed insights into the ground state properties by local probe techniques such as NMR and μSR are missing.

Herein, we present synthesis, magnetization, specific heat, muon spin relaxation, NMR, and inelastic neutron scattering results on the polycrystalline samples of the spin-orbit-driven frustrated 4$f$ magnet $Nd_3BWO_9$. Our thermodynamic results reveal the realization of a $J_{eff}$ =1/2 state in the lowest Kramers doublet, which is well separated from the excited states, as confirmed by our inelastic neutron scattering experiments. Thermodynamic results reveal the presence of a weak antiferromagnetic interaction between $Nd^{3+}$ moments ($J_{eff}$ = 1/2) at low temperatures. In addition, thermodynamic results point towards a magnetically ordered state below ~ 0.3 K and short-range spin correlations above 0.8 K. The muon spin relaxation results do not detect spin-freezing or magnetic ordering down to 34 mK possibly related to structural disorder and co-existence of static and fluctuating moments. The μSR relaxation rate is driven by an Orbach mechanism and dynamic electronic moments. NMR results are consistent with thermodynamic and μSR results. A fractionalized magnetization plateau has been observed in a magnetization isotherm recorded at low temperatures, suggesting that a complex magnetic phenomenon is active in this frustrated magnet. Our comprehensive results suggest that an intriguing field-induced magnetic phenomenon is at play in this spin-orbit-driven frustrated kagome magnet. The low temperature magnetism is driven by static and fluctuating magnetic moments, as well as short range spin correlations.

## II. EXPERIMENTAL DETAILS

Polycrystalline samples of $Nd_3BWO_9$ (henceforth NBWO) were synthesized following a standard solid-state reaction route. Stoichiometric amounts of $Nd_2O_3$ (Alfa Aesar, 99.999%), $H_3BO_3$ (Alfa Aesar, 98%), and $WO_3$ (Alfa Aesar, 99.998%) were taken as starting materials [44, 51]. Since $H_3BO_3$ is highly volatile, we have taken 5% of the excess to maintain proper stoichiometry in the sample. $Nd_2O_3$ was preheated at 900°C overnight to remove moisture and carbonates prior to use. The reagents were thoroughly mixed to achieve better homogeneity. The pellets were annealed in an alumina crucible in the temperature range of 700–900°C with several intermediate grindings. The non-magnetic analog $La_3BWO_9$ used for inelastic scattering experiment was prepared following the same method. For INS experiments, boron-11-enriched samples were synthesized. The x-ray diffraction (XRD) data were taken on the polycrystalline sample of Nd$_3$BWO$_9$ using a benchtop PANalytical diffractometer with CuK$_\alpha$ radiation ($\lambda$ = 1.541 Å) at room temperature.

The DC magnetization measurements on NBWO were performed in the temperature range of 1.8 ≤ $T$ ≤ 350 K using a Magnetic Properties Measurement System (MPMS3, Quantum Design). For measurements down to 400 mK, the MPMS3 was equipped with the iQuantum He3 setup. The DC magnetization data were obtained following zero-field and field-cooled protocols. The AC magnetization measurements were conducted in the temperature range from 1.8 to 60 K with 5–7 Oe AC excitation fields, up to 5 Tesla DC magnetic fields, and frequencies ranging from 10 Hz to 0.8 kHz using the AC option of the MPMS3. To measure the specific heat of the sample, a thermal relaxation method was employed using the specific heat option of the Physical Property Measurement System (PPMS, Quantum Design). For NBWO, the specific heat measurements were carried out in the temperature range of 0.06 to 4 K using a dilution refrigerator (DR) in zero field on PPMS from Quantum Design. The pulsed-field magnetization measurements up to 60 T were performed at the Dresden High Magnetic Field Laboratory (HLD), using a compensated pickup-coil magnetometer in a coaxial geometry with a pulse raising time of 7 ms. Each measurement of the sample was followed by recording the background without sample under identical conditions and its subsequent subtraction of the sample measurement, to avoid any background contributions [52]. The pulsed-field magnetization data were calibrated using magnetization data recorded in the static magnetic field on MPMS, Quantum Design. The field dependence of magnetization obtained by the two methods is in good agreement, confirming the accuracy of our experiments.

Inelastic neutron scattering measurements were carried out at the ISIS Neutron and Muon Source on the MARI time-of-flight spectrometer on Nd$_3$BWO$_9$ powder samples with an enriched 11-B isotope. The experiments were carried out using a close-cycle refrigerator (CCR) and a Fermi chopper with the Gd-slit package. We collected data at 5 K with a neutron incident energy of $E_i$ = 80 meV and a Fermi-chopper speed of 250 Hz. We also measured the phonon reference 11-B enriched compound La$_3$BWO$_9$ in identical conditions.

Muon spin relaxation ($\mu$SR) experiments were performed on GPS and FLAME spectrometers at the Swiss muon source at PSI in Villigen, Switzerland, in zero field and longitudinal field configurations. The measurements on the GPS instrument were performed down to 1.6 K on a pressed pellet 10 mm in diameter and 1.6 mm thick. The pellet was secured between two 25-μm Cu foils and fastened to a copper fork sample holder. Additionally, the FLAME spectrometer equipped with a Variox plus Kevinox dilution fridge insert was used for zero field (ZF) and longitudinal field (LF) μSR measurements in a wide temperature range between 0.034 and 300 K. The $\mu$SR data were analyzed to extract relevant parameters to shed insights into the ground state properties of the present material using the musrfit program [53].

$^{11}$B NMR ($I$ = 3/2) measurements were carried out using a home-built spectrometer and Oxford magnet with a fixed field of approximately 9.4 T corresponding to a Larmor frequency of 128.321 MHz. A flow cryostat was used to cover the temperature range between 4 and 300 K. A 133 mg

powder sample was used for these measurements. The NMR spectra were recorded by sweeping the frequency and combining individual spectra obtained via the standard Hahn-echo pulse sequence. The $^{11}$B spin-lattice relaxation was measured by employing the inversion recovery pulse sequence.

## III. RESULTS

### A. Structural details

The XRD data taken at room temperature confirms the phase purity of the samples studied in this work. To determine the structural parameters, Rietveld refinement of the XRD data was performed using GSAS software. The Rietveld refinement of the XRD data reveals no detectable atomic disorder in this material [Fig. 1(a-b)]. It reveals that NBWO crystallizes in a hexagonal structure with space group $P6_3$. In this material, $Nd^{3+}$ ions constitute a distorted kagome spin-lattice in the $ab$-plane that is stacked along the $c$-axis. The refinement parameters are consistent with earlier reports [51] and are summarized in Table I. It is observed that the $Nd^{3+}$ ions form $NdO_8$ dodecahedrons with oxygen ions which are distorted due to the slightly different Nd-O bond lengths, i.e., 2.7 Å, 3 Å, and 2.6 Å, resulting in a distorted kagome lattice in the $ab$-plane as shown in Fig. 1(a) that is consistent previous reports [44, 48, 49, 51]. The structure is distorted in the a$b$- plane as there are two inequivalent Nd triangles constituting the kagome network: one with an Nd-Nd bond length of ~4.2 Å and one with a bond length of ~4.9 Å. While the triangles are still corner-linked, the interstitial region is not a regular hexagon in shape, as can be seen in Fig 1(a), which suggests a complex exchange network in this frustrated magnet. The presence of a structural phase transition has been ruled out by the specific heat result presented in the next section.

Table I: The atomic parameters obtained from the Rietveld refinement of x-ray diffraction data of $Nd_3BWO_9$ recorded at room temperature. (Space group: P6$_3$, $a = b$ = 8.6997(1) Å, $c$ = 5.5006(1) Å, $\alpha = \beta$ = 90°, $\gamma$ =120° and $\chi^2$=1.7, R$_{wp}$ =4.78 %, R$_p$ =3.69 %, and R$_{exp}$ = 3.66 %).

| Atom | Wyckoff Pos. | x | y | z | Occupancy |
|---|---|---|---|---|---|
| Nd | 6$c$ | 0.3590(2) | 0.0832(1) | 0.229(2) | 1 |
| B | 2$a$ | 0.0000 | 0.0000 | 0.75 | 1 |
| W | 2$b$ | 0.3333 | 0.6666 | 0.255(2) | 1 |
| O1 | 6$c$ | 0.160(2) | 0.058(3) | 0.886(4) | 1 |
| O2 | 6$c$ | 0.199(2) | 0.444(2) | 0.046(3) | 1 |
| O3 | 6$c$ | 0.097(3) | 0.473(2) | 0.508(3) | 1 |

### B. Magnetization

The response of magnetization in the presence of an external applied magnetic field reveals interesting insights concerning exchange interactions, anisotropy, and spin correlations in NBWO. The temperature dependence of dc magnetic susceptibility taken in an external magnetic field of 0.1 T is shown in Fig. 1(c). The dc magnetic susceptibility shows neither signature of long-range magnetic order nor spin freezing down to 1.8 K. The inverse dc magnetic susceptibility data show two energy scales, wherein the high temperature $T > 150$ K behavior of inverse susceptibility corresponds to the thermal population of crystal-electric-field excitations [4]. The inverse magnetic susceptibility shows a change of slope below 100 K indicating repopulation of crystal electric field (CEF) levels. The Curie-Weiss fit of inverse magnetic susceptibility data below 10 K yields an effective moment $\mu_{\text{eff}} = 2.81\,(1)\,\mu_B$, which is small compared to the free ion moment value $\mu_{\text{eff}}^{\text{free}} = 3.62\,\mu_B$ of $Nd^{3+}$ ion ($4f^3$, $^4I_{9/2}$; $S=3/2$, $J=9/2$) indicating the realization of a low energy state with $J_{\text{eff}} = 1/2$ at low temperatures that might support strong quantum fluctuations. The reduced effective moment that is obtained from the fit of inverse magnetic susceptibility at low temperatures indicates that only some crystal field levels contribute to the susceptibility below 10 K, and thus there must be an energy gap between these low-lying levels and higher energy levels such that above 150 K we recover the full expected effective moment.

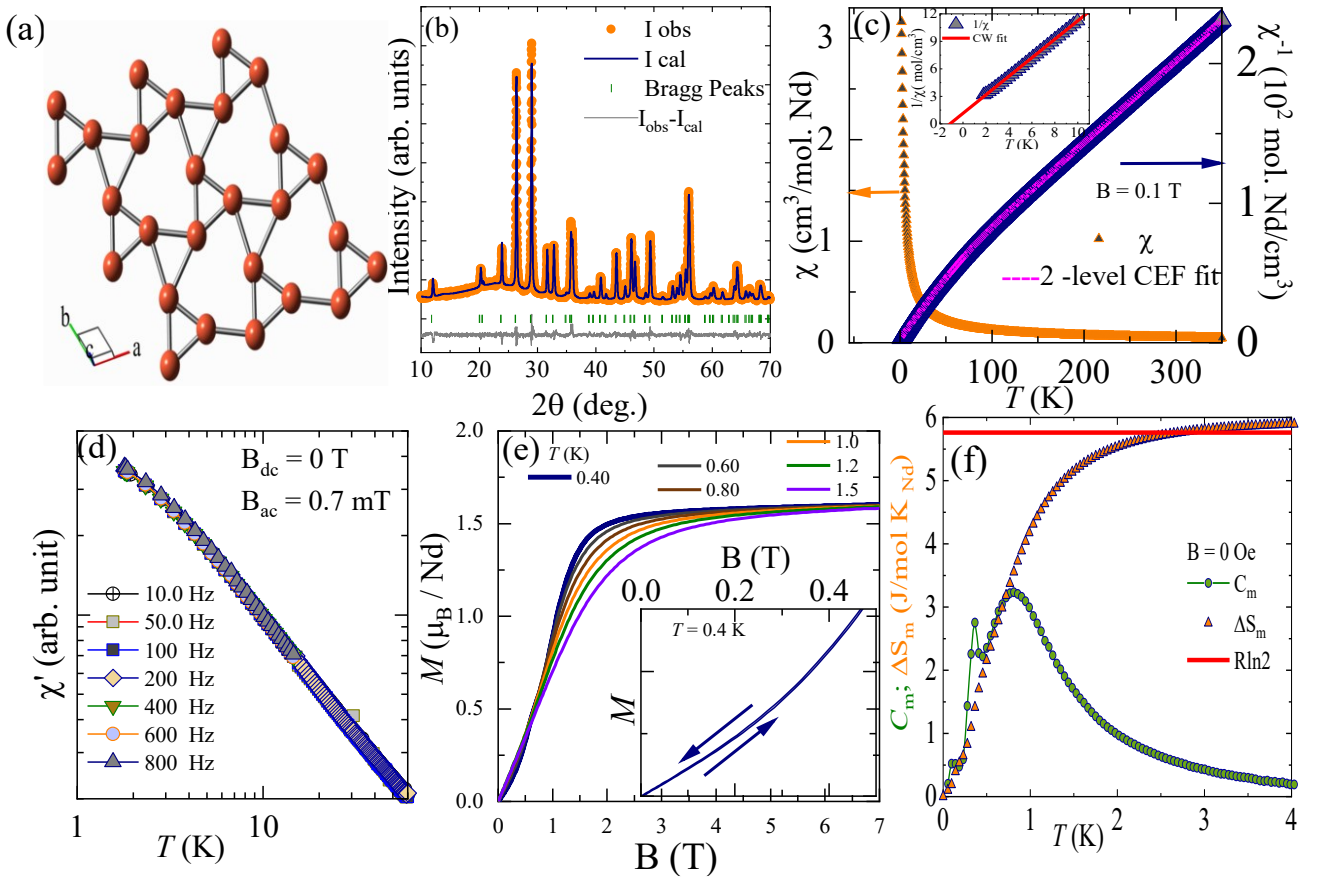

Fig.1 (a) Kagome lattice constituted by $Nd^{3+}$ ions in the $ab$-plane of $Nd_3BWO_9$. (b) Rietveld refinement of xrd data taken at room temperature. (c) The temperature dependence of magnetic susceptibility in 0.1 T (left) taken in ZFC mode and inverse magnetic susceptibility with Curie-Weiss and two level CEF fits as discussed in the text (right). The inset show the fit of $1/\chi$ vs. $T$ at low temperatures(d) AC susceptibility recorded at different frequencies in zero dc field and 0.7 mT ac field. (e) Magnetization isotherms at several temperatures. The inset shows the low-field magnetization isotherm at 0.4 K for better clarity. (f) The temperature dependence of magnetic specific heat after subtracting the lattice ($\sim T^3$) and nuclear ($\sim T^{-2}$) contributions in zero field at low temperature. The magnetic entropy in zero field that saturates to a value Rln2 reflecting a low energy $J_{eff}$ =1/2 Kramer doublet ground state at low temperatures.

The corresponding Curie-Weiss temperature, $\theta_{CW}$ = -1.2 (2) K represents the energy scale of exchange interaction between Nd$^{3+}$ ($J_{eff}$=1/2) moments at low temperatures. The crystal electric field affects the magnetic susceptibility at high temperatures in 4$f$ magnets. In NBWO, it splits the tenfold degenerate $J$ =9/2 multiplet into five Kramers doublets of Nd$^{3+}$ ion. The lowest Kramers doublet with $J_{eff}$ =1/2 at low temperatures is well separated from the excited states as revealed by our inelastic neutron scattering experiments as discussed in section B. The $J_{eff}$=1/2 lowest Kramers state mainly governs the low-temperature physics of this frustrated magnet. The mean field approximation, $J_{ex} = -\frac{3k_B \theta_{CW}}{zS(S+1)}$ (where $z$ = 4 and $S$=$J_{eff}$ =1/2 in the present case) yields an antiferromagnetic exchange interaction $J_{ex}/k_B$ = 1.2 (2) K. The presence of weak exchange interaction is typical of 4$f$ magnets owing to strong localization of rare-earth ions that is comparable with the dipolar interaction [36]. Our low temperature specific heat data reveal a phase transition around 0.3 K (see Fig. 1f), which is possibly due to an extra term in the magnetic Hamiltonian leading to a complex magnetic ordering phenomena. The frustration parameter, $f = \frac{|\theta_{CW}|}{T_N}$, quantifying the degree of frustration, is thus close to 5, suggesting a highly frustrated spin-lattice that enhances quantum fluctuations, leading to suppressed magnetic ordering at ∼0.3 K as discussed in the following section.

As depicted in Fig. 1(c), the two energy scales in the inverse magnetic susceptibility of the Nd$^{3+}$ ion could be reconciled using a simple two-level model-a simple approximation assuming that there are two levels [54] as described below.

$$\frac{1}{\chi(T)} = \frac{8(T - \theta_{CW})(1 + e^{-\Delta/k_B T})}{\mu_{1eff}^2 + \mu_{2eff}^2 e^{-\Delta/k_B T}}$$

where $\frac{\Delta}{k_B}$ = 215 (10) K is the gap between the ground state Kramers doublet and the first excited CEF level, $\mu_{1eff}$ = 3 (1) $\mu_B$ is the effective moment of the ground state, and $\mu_{2eff}$ = 4.2 (2) $\mu_B$ is the effective moment of the first excited state. The Curie-Weiss temperature $\theta_{CW}$ = −1.5(2) K is related to the exchange interaction between Nd$^{3+}$ moments in the Kramers ground state. The large value of CEF gap ($\frac{\Delta}{k_B}$), reflected as a change of slope of inverse susceptibility data below 100 K, suggests that the Kramers doublet ground state is well separated from the excited states, which is in reasonable agreement with that obtained from the inelastic neutron scattering experiment as discussed in the next section. As shown in Fig. 1(d), the ac susceptibility at different frequencies rules out a spin-glass state down to 1.8 K. The absence of hysteresis in the magnetization isotherm at 400 mK (see Fig. 1(e)) adds further credence to the claim. In the absence of specific heat on the non-magnetic analogue, the magnetic-specific heat obtained after subtracting the lattice ($C_L \sim T^3$) and nuclear-specific heat ($C_N \sim 1/T^2$) shows a broad maximum around 0.8 K, and on further lowering the temperature, an antiferromagnetic phase transition is observed at 0.3 K, as shown in Fig. 1(f). As presented in Fig. 1(f), the magnetic entropy extracted from the specific heat data saturates to a value that corresponds to Rln2 at low temperatures, suggesting the realization of a low-energy $J_{eff}=1/2$ state in the ground state Kramers doublet, which is in agreement with magnetic susceptibility.

**C. Crystal Electric Field Spectrum**

The magnetism of the frustrated kagome magnet $Nd_3BWO_9$ is innately connected to the crystal electric field scheme of $Nd^{3+}$ ion. To investigate the crystal field excitations, and anisotropy and to probe the energy and wave vector dependence of the dynamic structure factor in $Nd_3BWO_9$, we have performed INS experiments on MARI spectrometer. Fig.2 (a) shows the Q-integrated (0 to 4 Å$^{-1}$) INS spectra from low wave vector Q in $Nd_3BWO_9$. It shows magnetic scattering estimated after subtracting the phonon using the nonmagnetic $La_3BWO_9$ data (see appendix Fig. S1) [55]. We have observed four well defined CEF excitations from the ground state CEF doublet of $Nd^{3+}$ (ground multiplet $J = 9/2$, which spits into 5 CEF doublets in the $C_1$ point symmetry of $Nd^{3+}$ ions in the paramagnetic state), in $Nd_3BWO_9$ (Fig. 2b), which indicate the localized nature of 4f-electron in this kagome magnet. The solid line shows the fit based on the CEF analysis and the values of CEF parameters obtained are given in the Appendix-A, Tables II, III and IV.

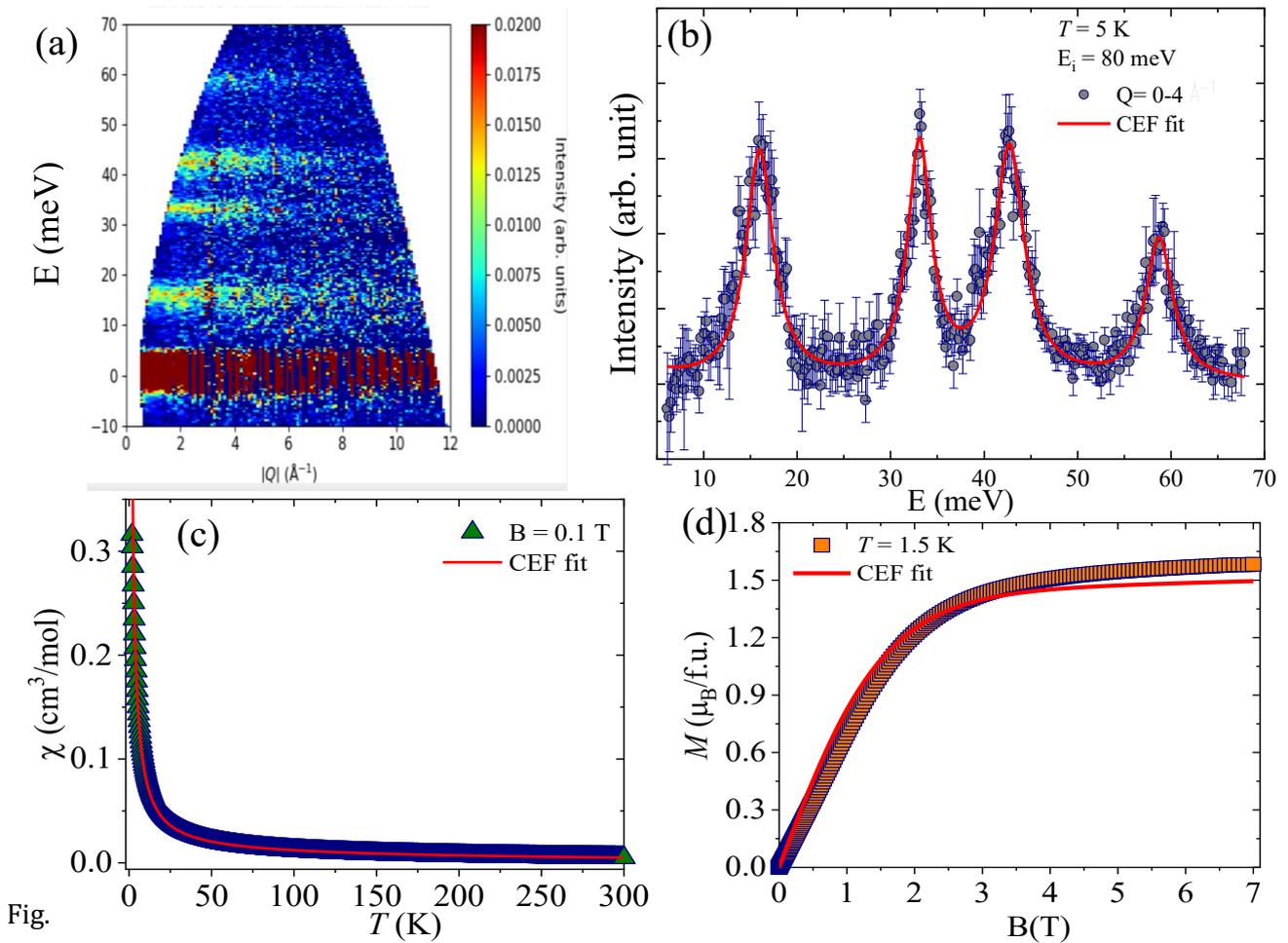

Fig.

2 (a) Inelastic neutron scattering intensity of Nd$_3$BWO$_9$ obtained after subtraction of phonon contribution using the non-magnetic reference La$_3$BWO$_9$ at 5 K for the incident energy E$_i$=80 meV (b) Q integrated (0 to 4 Å$^{-1}$) E vs. Intensity-cut from the INS data taken with E$_i$=80 meV at 5 K with fit to the CEF model shown by the solid line. (c) The temperature dependence magnetic susceptibility data on the polycrystalline sample and the solid line is the calculated magnetic susceptibility using the CEF parameters as provided in the appendix. (d) The field dependence of magnetization isotherm at 1.5 K and the solid line is the calculated magnetization isotherm at 1.5 K with relevant CEF parameters.

The CEF parameters were estimated by fitting the INS spectra and magnetization data taken on the polycrystalline sample together (see Appendix for more details). The calculated magnetic susceptibility along with the experimental susceptibility data are shown in Fig. 2c. We also simulated the magnetization isotherms at 1.5 K using the CEF parameters and the results are plotted in Fig.2d, which also agree reasonably well with the experimental *M* vs. *H* results on NBWO [49]. The value of the CEF gap 186(8) K suggests that the lowest Kramers doublet is well separated from the excited states confirming the thermodynamics and µSR results. Thus, the ground state physics of this kagome antiferromagnet could be mapped by the low energy state $J_{eff}$ =1/2 corroborating thermodynamic results. The $J_{eff}$ =1/2 state arises due to strong spin-orbit coupling and crystal electric field likely to induce strong quantum fluctuations at low temperatures and likely to dominate ground state of this frustrated magnet.

**D. AC susceptibility and specific heat**

AC susceptibility is a sensitive probe to track ground state properties and spin dynamics in frustrated magnets. AC susceptibility measurements were performed in zero DC field and in AC excitation magnetic fields in a wide frequency range (10Hz **to 800 Hz**) to delineate spin dynamics and shed insights into field-induced effects in this kagome antiferromagnet. As shown in Fig. 1d, the real part of AC susceptibility ($\chi'_{ac}$) recorded in an AC field of 7 Oe increases monotonically upon decreasing temperature without any signature of magnetic ordering down to 1.8 K, which is consistent with the DC magnetic susceptibility.

The frequency-independent behavior of AC susceptibility down to 18 K suggests that the characteristic fluctuation rate of Nd$^{3+}$ moments is beyond the time scale of AC susceptibility experiment. Fig. 3 (a-b) depicts the AC susceptibility taken in several DC fields with an AC excitation field of 7 Oe at a frequency 10 Hz. The real part of AC susceptibility ($\chi'_{ac}$) in DC magnetic fields shows a field-induced behavior at around 7 K and a noticeable change of slope with strong DC fields (Fig. 3b). Notably, the imaginary part of the AC susceptibility ($\chi''_{ac}$) exhibits a broad maximum ~4.3 K in DC magnetic fields of 0.5 T, and the position of the maximum shifts towards higher temperature upon increasing magnetic field suggesting the emergence of a field induced gap. Contrastingly, the amplitude of $\chi''_{ac}$ starts decreasing with a DC magnetic field ≥ 2 T [56]. The magnetic origin of the broad maximum in specific heat at ~0.8 K in zero field is confirmed by Fisher's specific heat $\left(\frac{d(\chi T)}{dT}\right)$ calculated from the DC susceptibility recorded down to 0.4 K (see Fig.3(c)). It is worth noting that this maximum is independent of the measurement protocol (ZFC-FC) and is insensitive to magnetic field up to 0.1 T. However, the 0.8 K maximum flattens with magnetic field above 0.4 T and shifts towards lower temperature and completely disappears with a field ≥ 0.9 T. Interestingly, the magnetic specific heat shows an anomaly at ~0.4 K in a field of 0.9 T, which indicates that a field-induced phenomenon is at play in this kagome antiferromagnet. This is indeed consistent with the presence of anomaly in isothermal magnetization recorded at low temperatures as discussed in the next paragraph. In order to single out potential magnetic ordering, we have performed magnetization and specific heat measurements also in the sub-Kelvin temperature range given weak interaction energy scale in this localized 4$f$ magnet.

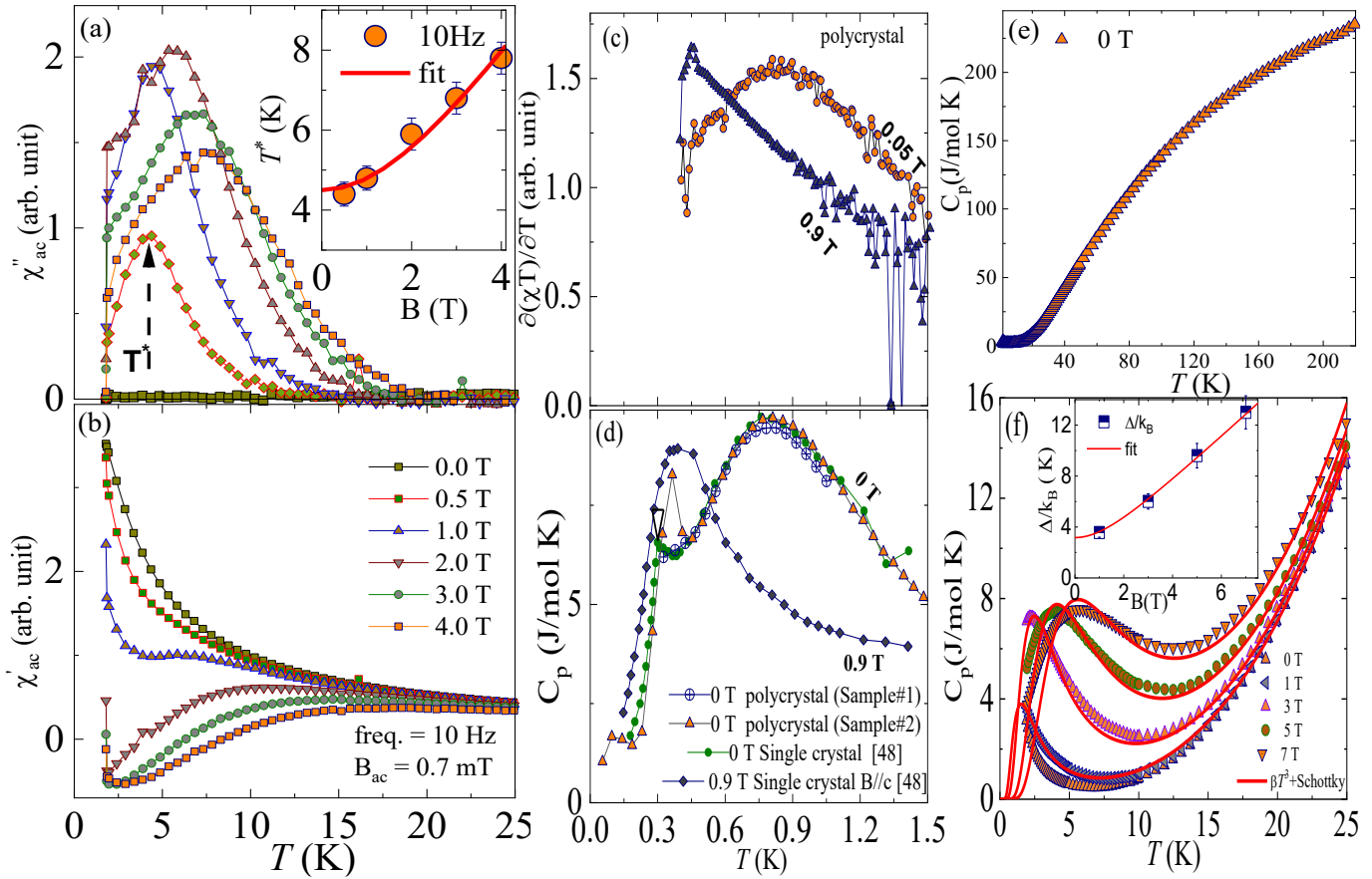

Fig.3. (a) The temperature dependence of the imaginary part of the ac susceptibility and the inset shows the variation of broad maxima ($T^*$) in the imaginary part of AC susceptibility with DC magnetic field. The solid line is a fit as discussed in the text (b) The temperature dependence of the real part of the ac susceptibility in different DC fields taken at 10 Hz and ac field 0.7 mT. (c) The temperature dependence of Fisher specific heat determined from dc susceptibility measured in 0.05 T and 0.9 T (d) The temperature dependence of specific heat in the polycrystalline samples of $Nd_3BWO_9$ that is compared with the single crystal data adapted from ref. [48]. (e) The temperature dependence of specific heat in zero field in the temperature range 1.8 ≤ $T$ ≤ 220 K rules out the presence of any structural phase transition. (f) The temperature dependence of specific heat taken in different applied fields at low temperature. The solid line is a fit to Schottky model plus $T^3$ term accounting for the phonon contributions as discussed in the text and the inset shows the field dependence of Zeeman gap with a fit as discussed in the next.

Fig. 3d depicts the temperature dependence of specific heat in sub-Kelvin temperature taken in zero field that is compared with that taken on the single crystals adapted from ref. [48]. The specific heat shows a sharp anomaly at ~0.3 K typical of long-range magnetic ordering. The anomaly at 0.3 K is sensitive to applied magnetic field and it smears out with the magnetic field. The specific heat shows a broad peak around 0.4 K in an external magnetic field of 0.9 T. The appearance of a broad maximum in specific heat around 0.8 K in the absence of a magnetic field and the lower value of magnetic entropy release ~2.5 J/mol. K at the ordering temperature ~0.3 K compared to the expected value 5.76 J/mol. K in zero field are characteristic features of short-range spin correlations. Furthermore, the position of broad maximum in magnetic specific heat is in close agreement with that expected at $T/J_{ex}$~0.6 for S=1/2 kagome quantum magnet add further credence to the existence of short-range spin correlations [57, 58]. The specific heat results indicate the co-existence of long range and short-range order in this kagome material. The specific heat in the temperature range from 1.8 K to 220 K (see Fig. 3(e)) don't show any anomalies that suggest the absence of a structural phase transition in the temperature range of investigation in this material. Fig. 3(f) shows the temperature dependence of specific cheat in different external applied magnetic fields. As shown in Fig. 3(f), the broad maximum shifts in a Schottky-like manner toward higher temperature upon increasing the magnetic field. In order to get an estimate of gap value owing to the Zeeman splitting of the lowest Kramers doublet in the presence of applied magnetic field, the specific heat data in magnetic field is fitted with two level Schottky model and $T^3$ term due to phonon contribution; $C_p = \beta T^3 + C_{Sch}$ with $C_{Sch} = fR\left(\frac{\Delta}{k_B T}\right)^2 \frac{e^{\frac{\Delta}{k_B T}}}{\left(1+e^{\frac{\Delta}{k_B T}}\right)^2}$, where $f$ represents the fraction of paramagnetic $Nd^{3+}$ ions participating in the splitting of the ground state doublet, R is the universal gas constant, $\Delta$ is the Zeeman gap, $k_B$ is the Boltzmann constant. The field-induced gap that scaled with magnetic field is ascribed to the Zeeman splitting of the lowest Kramers doublet $J_{eff}$=1/2 at low temperatures (inset of Fig. 3f). To estimate the value of the field induced gap, the $T^*$ vs H obtained from ac susceptibility (inset of Fig. 3a) and $\Delta$ vs. H obtained from the Schottky model of the specific heat (inset of Fig. 3f), were fitted with zero-field gap plus a Zeeman term in quadrature $\Delta(H) = \sqrt{\Delta_0^2 + (g\mu_B H)^2}$ [50]. The obtained Landé g-factor is in close agreement with that found from the of magnetization results.

To shed more insight into the effect of magnetic field on the low-temperature magnetic phase, we have performed magnetization measurements down to 400 mK and in magnetic fields up to 60 T. Fig. 4(a) shows the isothermal magnetization recorded at 400 mK in $Nd_3BWO_9$. The absence of hysteresis rules out any secondary phases or spin glass behavior in this frustrated kagome material. The magnetization increases with field up to 0.4 T followed by a quasi-plateau wherein the orientation of the magnetic moment changes much less with the external applied magnetic field. This is consistent with the 1/3 $M_{sat}$ magnetization plateau and eventually saturates to a value of $M_{sat}$ = 1.5 $\mu_B$/Nd, as shown in Fig. 4(a), in line with what was found in the single crystals [48,49]. The observed features in the magnetization isotherm at 400 mK imply a complex magnetic structure wherein phenomenologically, two spins in a triangle of the host kagome lattice might align with the external magnetic field while one spin in each triangle points opposite to the applied magnetic field, known as the uud phase. This exotic magnetic phase is associated with a complex magnetic phenomenon due to the fact that quantum fluctuations lift ground state degeneracy and adopt specific spin configurations at low temperatures in frustrated magnets. [1, 59-63]. Upon further increase of the magnetic field, the magnetization isotherm at 400 mK saturates completely to a constant value ∼1.5 $\mu_B$/ $Nd^{3+}$ in an external magnetic field of 4 T. The lower value of saturation magnetization compared to the expected value, which indicates the role of magnetic anisotropy or frustration induced quantum fluctuations as has been observed in several frustrated magnets [48,49, 64, 65]. As shown in Fig.4(b), the derivative of isothermal magnetization $dM/dB$ at 400 mK reveals many interesting features, namely, a shoulder-like anomaly, two peaks corresponding to critical fields $B_{c1}$, $B_{c2}$ related to the beginning and end of magnetization plateau and a minimum at $B_{min}$ at the center of the magnetization plateau [59-63]. The shoulder-like anomaly shifts to the lower magnetic field side and vanishes upon increasing the temperature above 0.5 K. The anomalies at $B_{c1}$ and $B_{min}$ slightly shift to the higher field side and also eventually disappear at $T > 0.5$ K. The peak at $B_{c2}$ shifts to the lower magnetic field side upon increasing the temperature and disappears within a broad hump at $T \geq 0.8$ K. This intriguing low-temperature behavior of the magnetization suggests that a complex field-induced quantum phenomenon such as metamagnetic mechanism is at play in this spin-orbit driven kagome antiferromagnet. Further experiments on the high quality single crystals may shed interesting insights in this context.

Fig. 4(c-d) represent a comparative account of the magnetization isotherms taken at different temperatures using a static field and pulsed magnetic field up to 60 T. The magnetization isotherm at low temperature $T \sim 540$ mK saturates in the low field regime and increases with field in the high field limit (Fig. 4(c)). The high-field magnetization isotherm at 11.5 K shows non-linear behavior over a wide field range, as shown in Fig. 4(c-d). The linear term in the high-field magnetization isotherm indicates the presence of short-range spin correlations that are consistent with the enhancement of temperature dependence of magnetic susceptibility and NMR shift at low temperatures, as discussed in subsequent sections.

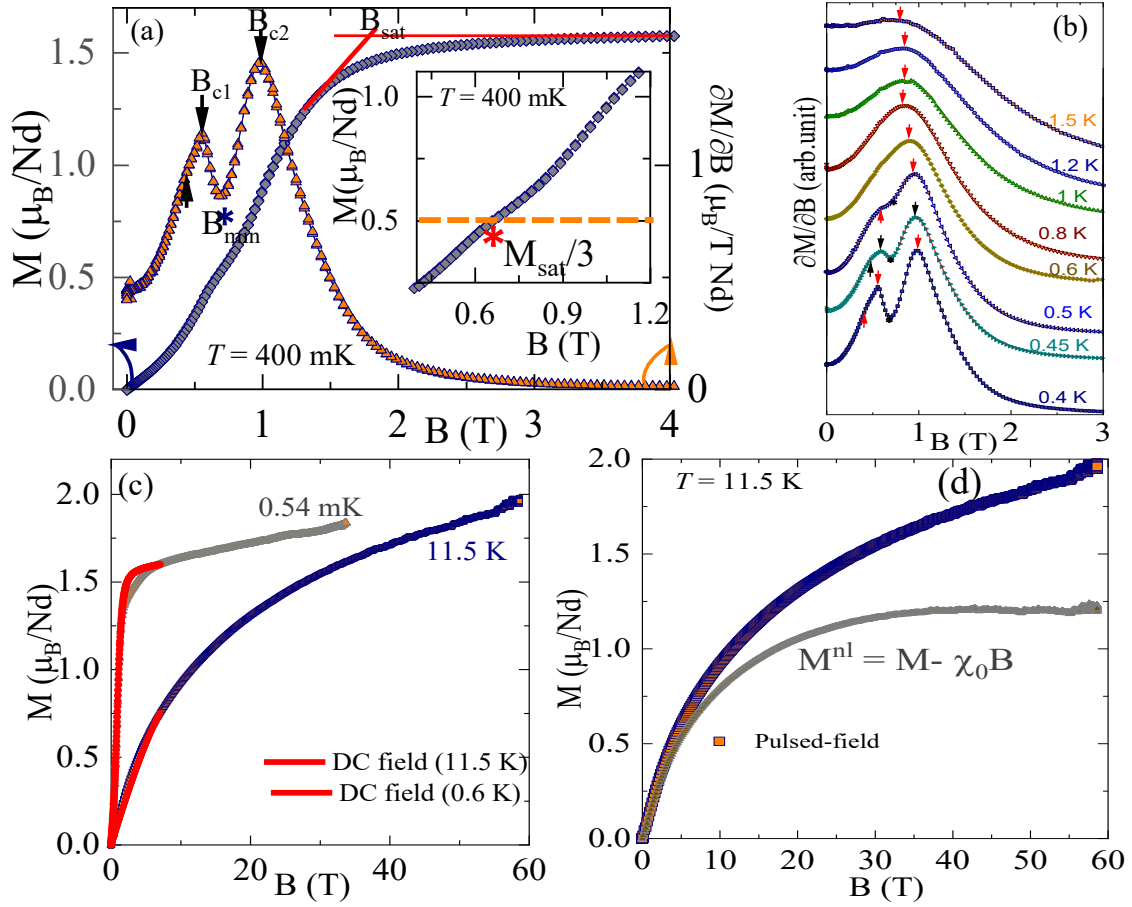

Fig. 4. (a) Magnetization isotherm recorded at 400 mK (left). The derivative of magnetization ($\partial M/\partial B$) with respect to magnetic field (right) is also shown to track the critical fields associated with some field induced phases at low temperature. The inset shows the magnetization plateau at 400 mK. (b) The derivative of magnetization isotherms ($\partial M/\partial B$) taken at different temperature to identify the evolution of field induced phases in an external applied magnetic field. (c) Magnetization isotherm measured up to 60 T at different temperatures using pulsed magnetic field. The pulsed field data are compared with that taken in a static magnetic field. (d) Magnetization isotherm at 11.5 K taken using pulsed high magnetic field up to 60 T, where $M^{nl}$ corresponds to non-linear magnetization as discussed in the text.

Remarkably, the specific heat, $C_p$ and Fisher specific heat derived from DC magnetic susceptibility, $\left(\frac{d(\chi T)}{dT}\right)$, data show a broad maximum around 0.8 K that is indicating a cross-over from a paramagnetic phase to a field-polarized state consistent with our magnetization experiments conducted at low temperatures. The $Nd_3BWO_9$ frustrated kagome material behaves like an unconventional paramagnet above 0.8 K.

## E. Muon spin relaxation

In order to shed additional light to the magnetism of NBWO and track the local field distribution in real space as well as spin fluctuations, zero field (ZF) and longitudinal field (LF) $\mu$SR experiments were conducted over a wide temperature and field range.

### E.1 ZF measurements

The $\mu$SR measurements were first performed in zero field (ZF) with the initial spin direction anti-parallel to the muon momentum. Fig. 5a depicts the ZF muon asymmetry at the lowest temperature of 34 mK. The absence of oscillations in the ZF muon asymmetry curve and the vanishing asymmetry at long times, i.e. the absence of the 1/3 tail, rule out static local magnetic fields at the muon stopping sites even at the lowest accessible temperature. The $\mu$SR asymmetry relaxes extremely quickly (note the logarithmic scale in Fig. 5a), however, it is composed of two contributions, one faster and one slower decaying. The attempts to fit the data with the stretched exponent relaxation function were unsuccessful, a model with two simple exponentially decaying components works much better. Thus, we use the model $A(t) = A_0 \big( f \cdot e^{-\lambda_{\text{fast}} t} + (1 - f) \cdot e^{-\lambda_{\text{slow}} t} \big)$, with the fraction of the fast component $f = 0.86(1)$ fixed from low-temperature data. The presence of two components is likely due to muons stopping at two magnetically inequivalent muon stopping sites in the host spin-lattice.

The temperature dependence of the muon spin relaxation rate of the dominant fast component in zero field is shown in Fig. 5(b). The muon relaxation rate increases by several orders of magnitude on cooling the sample from 300 to 20 K, where a relaxation plateau is established. The strong temperature dependence of the ZF relaxation rate is consistent with an Orbach relaxation mechanism, associated with crystal-field fluctuations, as observed in several frustrated magnets including NdTa$_7$O$_{19}$ [2, 66].

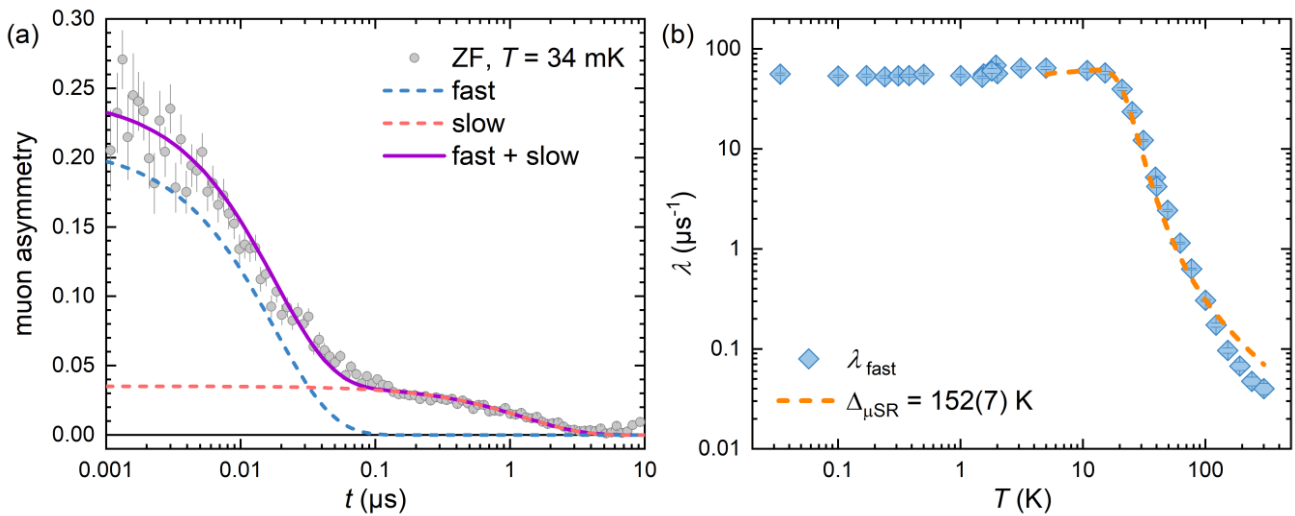

Fig. 5 (a) The ZF muon asymmetry dataset (points) at the base temperature with the two-component exponential fit (lines), as further explained in the text. (b) The temperature dependence of the ZF muon relaxation rate of the fast-relaxing dominant component. The dashed line depicts the model combining the temperature-independent and the Orbach relaxation mechanisms as explained in the text.

Assuming the fast fluctuation regime, the muon relaxation rate is inversely proportional to the frequency of spin fluctuations, $\lambda^{-1} \propto \nu_e$, which is described as a sum of two independent mechanisms,

$$\nu_e(T) = \nu_0 + \frac{\nu_1}{e^{\Delta_{\mu SR}/k_B T} - 1}. \quad (1)$$

Here, the temperature-dependent term is the Orbach term that dominates the relaxation rate above 20 K. The constant term that corresponds to the relaxation rate plateau is associated with the muon spin relaxation in the ground-state Kramers doublet. Such behavior of the relaxation rate suggests that at temperatures below 20 K effective spin-1/2 degrees of freedom emerge in the ground state, as the excited crystal-field states do not contribute anymore to the muon spin relaxation. The relaxation rate was modelled with Eq. (1) after combining the ZF and the LF measurements, which allowed for the determination of the coupling constant between $\lambda^{-1}$ and $\nu_e$, as shown in the next section. The fraction of fast component is 86% while the fraction of slow component is just 14 %. The relaxation rate of the slow component follows a behavior akin to that of the fast component, however, the relaxation rate is one orders of magnitude lower than that of the fast component (not shown here).

### E.2 LF measurements

To gain a more quantitative insight into the fluctuations of the $Nd^{3+}$ magnetic moments, we performed LF measurements in a wide range of temperatures in three magnetic fields between 1.5 and 3.5 T.

The LF muon asymmetry curves were fitted with the same two-component exponential model as for ZF with the fraction of the two components fixed to $f = 0.86$, and the resulting fast-relaxing component is shown in Fig. 5a. In the fast fluctuation regime, the field and the temperature dependence of the relaxation rate is described by the Redfield model [67]

$$\frac{\lambda(T)}{\chi(T)T} = A_{\mu SR} \frac{\nu_e(T)}{\nu_e(T)^2 + \omega_L^2}. \quad (2)$$

Since the electron spin fluctuation frequency $\nu_e(T)$ monotonically increases with increasing temperature due to the Orbach process, a maximum appears in the muon spin relaxation rate at a temperature where $\nu_e$ crosses the Larmor frequency $\omega_L = \gamma_\mu B_0$ for each LF (Fig. 6a). From the maximum, we can extract the value of the coupling constant $A_{\mu SR}$ for the three applied fields. It is found to be field independent within the error bars, so we use its average value $A_{\mu SR} = 33.2 \cdot 10^3$ mol/(K·cm³·s²) to derive the temperature dependence of $\nu_e(T)$ for each field including the ZF measurement. This rate is found to be field independent in the investigated range between 0 and 3.5 T (Fig. 6b).

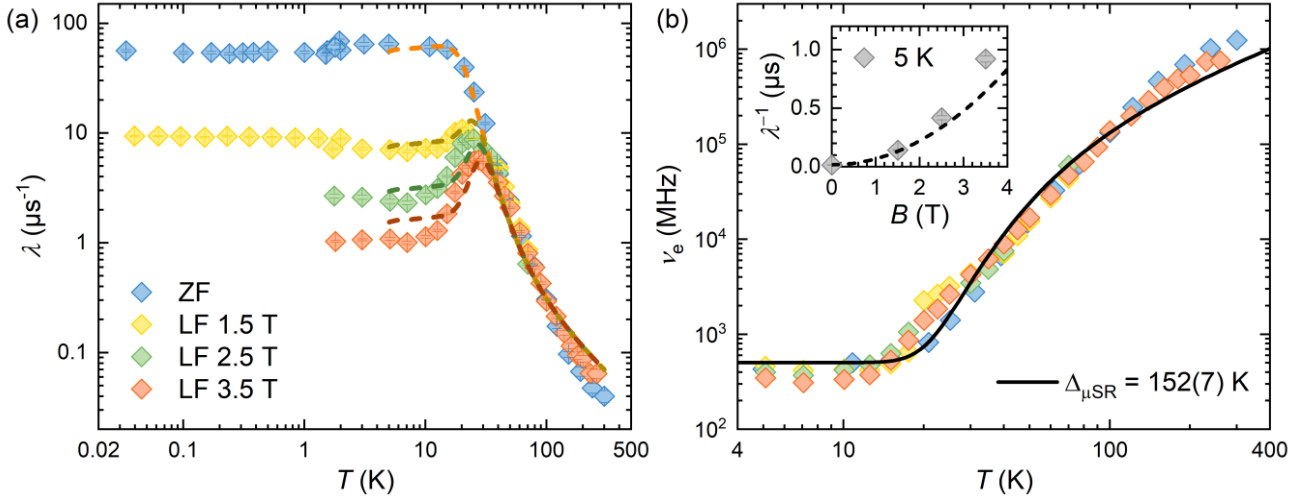

Fig. 6 (a) The temperature dependence of the muon spin relaxation rate of the dominant fast component and (b) the corresponding spin fluctuation rate $\nu_e$ (see text) at ZF and at several LFs. The inset shows the field dependence of the muon spin relaxation rate at 5 K. The solid line in panel (b) is a fit of $\nu_e$ in all magnetic fields to Eq. (1). The obtained model is then used to calculate the relaxation rates with Eq. (2), which are shown with the dashed lines in panel (a) and the inset.

The simultaneous fit of the four $\nu_e(T)$ datasets with a single-gap Orbach model plus a constant (Eq. (1), three free parameters) yields a low-temperature plateau of $\nu_0 = 504(43)$ MHz and the gap of $\Delta_{\mu SR}/k_B = 152(7)$ K that roughly matches the gap between the ground state and the first excited CEF level $\Delta/k_B = 186(8)$ K that was estimated via the INS, again revealing that the lowest Kramers doublet is well separated from the excited Kramers doublets. A similar scenario was observed in several rare-earth-based frustrated magnets [4, 66].

Finally, with the model for $\nu_e(T)$ and a coupling constant $A_{\mu SR}$ we can use Eq. (2) to reconstruct the temperature dependence of the muon spin relaxation data in all fields including the ZF (Fig. 6a), as well as its field dependence at 5 K (inset in Fig. 6b). Since we use a fairly simple model, the results at high temperatures are not described perfectly, likely due to higher CEF levels affecting the spin fluctuations, and there is some deviation at high fields as well. Nevertheless, the model is remarkably successful in describing the main features of the data in all fields. At low temperatures, the spin fluctuating frequency $\nu_e = 504(43)$ MHz can be compared to the width of the fluctuating fields distribution at the muon stopping site of $B_{loc} = \gamma_\mu^{-1}\sqrt{A_{\mu SR}\chi(T)T/2} = 0.14(4)$ T, which corresponds to $\gamma_\mu B_{loc} = 119(36)$ MHz. The assumption of the fast fluctuation regime ($\nu_e > \gamma_\mu B_{loc}$) was therefore justified even at low temperatures. It may be noted that μSR does not detect an anomaly at 0.3 K, unlike thermodynamic experiments. The absence of oscillations in the μSR does not necessarily indicate the absence of long-range order. If slow magnetic fluctuations (in the MHz) are present together with static magnetism, the muon spin relaxation might be dominated by the magnetic fluctuations at low temperatures. Therefore, our μSR study invokes the search for a persistent spin dynamic that coexists with the magnetic order indicated by the macroscopic techniques. Whether the emergence of exotic fractional excitations in frustrated magnets could lead to such a scenario is not clear at present, but it offers a new direction to carry out detailed investigations on the single crystals of NBWO in the future [68, 69].

**F. Nuclear magnetic resonance**

Nuclear magnetic resonance (NMR) is a powerful technique for providing microscopic insights into the intrinsic magnetic susceptibility and low energy spin dynamics. In NBWO, there is only one crystallographic site for boron, with the Wyckoff position 2a. The $^{11}$B NMR spectra shown in Fig. 7a consist of the central line and two satellite lines, as typical for $I = 3/2$ nuclei. The quadrupole frequency $\nu_Q \approx 1.3$ MHz is determined from the distance between the two satellite peaks. The spectra get increasingly broadened with lowering temperatures, the change in the spectrum is especially large between 45 K and 40 K. In the intermediate temperature range 10 ≤ $T$ ≤ 40 K, we are not able to measure the NMR spectra due to extremely short spin-spin relaxation times $T_2$ (Fig. 7b). Such NMR wipe-out was observed before in another Nd-based kagome system, the neodymium langasite [70, 71]. However, the NMR signal reappears below 10 K due to the increase of the $^{11}$B spin-spin relaxation time, in line with the slowing down of spin dynamics that observed in µSR (Fig. 6b).

On cooling, the NMR spectra broaden and shift to lower frequencies. The $^{11}$B NMR shift comprises of temperature dependent spin shift and temperature independent orbital/chemical shift $K_0 = 0.13(1)\%$; $K(T) = K_{\rm spin} + K_0$. Here, $K_{\rm spin}$ scales with bulk magnetic susceptibility, i.e., $K_{spin}(T) = A_{hf}\chi(T)$, in the whole temperature range between 300 and 40 K where the position of the central line can be unambiguously determined (Fig. 7c). The fit yields the hyperfine coupling constant between nuclear and electronic spins, $A_{hf} = -0.178(4)$ T/$\mu_B$.

In order to probe dynamic susceptibility associated with low-energy spin fluctuations, we also performed $^{11}$B nuclear spin-lattice ($T_1$) relaxation time measurements [72]. The temperature dependence of the spin-lattice relaxation rate $1/T_1$ is shown in Fig. 7b. Above 50 K, it was measured at the maximum of the central line (stars in Fig. 7a), for which the inversion recovery curve of the longitudinal nuclear magnetization $M$ was fitted by a multi-exponential recovery curve valid for magnetic relaxation in the case of $I = 3/2$ nucleus,

$$M_z(t) = M_0 \left[1 - (1+\alpha)\left\{0.1 \exp\left(-\left(\frac{t}{T_1}\right)^\beta\right) + 0.9 \exp\left(-\left(\frac{6t}{T_1}\right)^\beta\right)\right\}\right].$$

Here $\alpha$ is the inversion factor and $\beta$ is the stretching exponent. The stretched exponential behaviour (i.e., $\beta < 1$ in Fig. 7d) suggests a distribution of relaxation times in the host spin-lattice. Below 50 K, $1/T_1$ was measured on the right satellite peak (triangles in Fig. 7a) due to better signal compared to the central line. Here, the magnetization was modelled with an expression appropriate for the satellite transition,

$$M_z(t) = M_0 \left[1 - (1+\alpha)\left\{0.1 \exp\left(-\left(\frac{t}{T_1}\right)^\beta\right) + 0.5 \exp\left(-\left(\frac{3t}{T_1}\right)^\beta\right) + 0.4 \exp\left(-\left(\frac{6t}{T_1}\right)^\beta\right)\right\}\right].$$

In order to verify that the $1/T_1$ values from different parts of the spectra can be directly compared, we measured $1/T_1$ at both the central and the satellite transition at 95 K, where practically the same value was found (Fig. 7d).

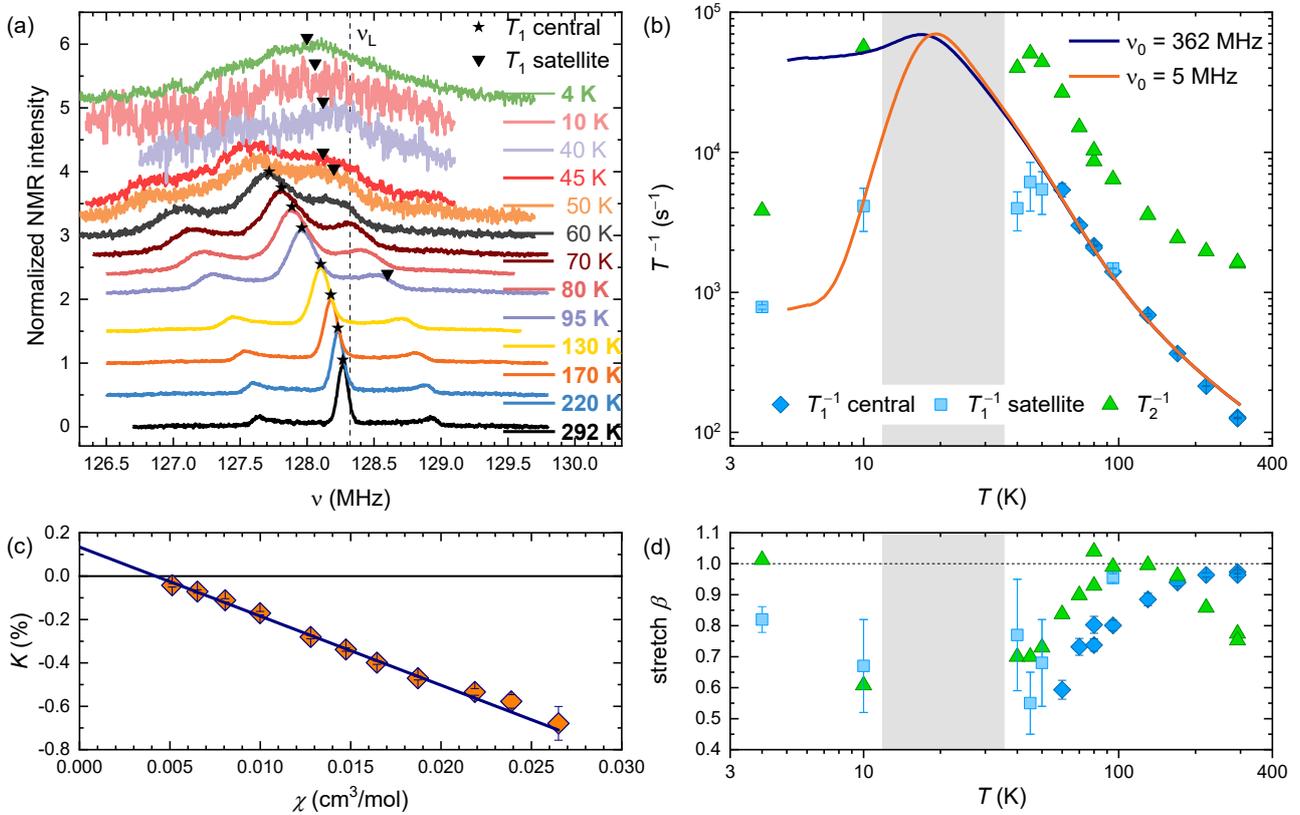

Fig.7 (a) The temperature evolution of $^{11}$B NMR spectra recorded at 9.4 T. The dashed line corresponds to the Larmor frequency. Stars and triangles indicate the positions where $T_1$ measurements were conducted at a particular temperature. The scaling of the central line with bulk susceptibility is shown in (c) together with the linear fit (solid line). (b) The temperature dependence of the spin-lattice ($1/T_1$) and the spin-spin ($1/T_2$) relation rate and (d) the corresponding stretch exponents. The solid lines in (b) correspond to the fits of the BPP model to the $1/T_1$ dataset, as explained in the text.

The temperature dependence of $1/T_1$ matches the temperature dependence of the muon spin relaxation rate $\lambda$ (Fig. 6a). In fact, we can use the same model as in the μSR analysis,

$$\frac{1}{T_1 T \chi(T)} = A_{NMR} \frac{\nu_e(T)}{\nu_e(T)^2 + \omega_L^2},$$

which is in NMR known as the Bloembergen, Purcell and Pound (BPP) theory [72, 73]. In our NMR experiment, we cannot measure the spin-lattice relaxation at the maximum value (where $\nu_e(T) = \omega_L$), because of the wipe-out, therefore the coupling constant $A_{NMR}$ cannot be determined from the experiment, as it was possible for $A_{\mu SR}$. Therefore, we use the same two-gap model with the parameters from the μSR modelling, with the rescaled $A_{NMR} = 3500\, A_{\mu SR}$ as the only free parameter. The resulting curve fits well with the experimental $1/T_1$ values in the temperature range between 300 and ~50 K (blue curve in Fig. 7b). However, at low temperatures this model fails to account for the experiment results. A better agreement is obtained (red curve in Fig. 7b) if the plateau value of the fluctuation frequency is also a free parameter when the fit yields $v_0$ = 5(1) MHz. The decrease of the low-temperature fluctuation rate from $v_0$ = 457(43) MHz, as determined from μSR, might be a result of a much stronger magnetic field applied in the NMR experiments (9.4 T) that should sufficiently polarize the spins at low temperatures. Alternatively, in the case of spin correlations, different local probes can filter spin fluctuations differently because of the site dependence of the form factor.

## IV. DISCUSSION

Competing magnetic interactions and spin-orbit driven magnetic anisotropy conspire with quantum fluctuations in driving intriguing quantum phenomena in the present kagome antiferromagnet $Nd_3BWO_9$. The present kagome material is a distorted quantum magnet with corner-shared triangular motifs wherein inequivalent Nd-Nd bond distances and Nd-O-Nd bond angles most likely led to inequivalent exchange couplings in the host spin-lattice. The kagome planes are separated by a distance comparable to the intraplane bond distance. In this material with $Nd^{3+}$ ($^4I_{9/2}$) ions, the crystal-field states comprise of five Kramers doublets, and the ground state with $J_{eff}$ =1/2 is well separated from the first excited state by a gap of 186 (8) K, as evidenced by our inelastic neutron scattering experiments. The low-temperature physics of this Kagome antiferromagnet is governed by the exchange interaction between $J_{eff}$ =1/2 moments. In the mean-field approximation, the antiferromagnetic exchange interaction between $Nd^{3+}$ moments determined from the Curie-Weiss temperature obtained from the fit of magnetic susceptibility at low temperatures is $J_{ex}$ =1.2 K (2). The presence of such a weak exchange interaction is typical of 4$f$ magnetic moments in view of the strongly localized nature of 4$f$ orbitals [2–4]. Specific heat measurements show a λ-like sharp anomaly around 0.3 K, implying an antiferromagnetic phase transition that is associated with the presence of finite interplane interaction between $J_{eff}$ =1/2 spins [49]. The low-temperature magnetization isotherm in the magnetically ordered state exhibits a plateau at 1/3 of the saturation magnetization value in the presence of an external magnetic field of ~0.5$J_{ex}$. This unconventional behavior of low temperature magnetization isotherm in this frustrated kagome antiferromagnet is most likely related to a complex magnetic structure comprising up-up-down spin configurations of $J_{eff}$ = ½ spins in the host spin-lattice [59-63]. This type of magnetic behavior has been encountered in several frustrated 3$d$ and 4$f$ magnets with spontaneous breaking of translational symmetry [59-63]. In the $S$=1/2 Heisenberg model on a kagome lattice wherein spins interact via exchange coupling $J_{ex}$, the 1/3 magnetization plateau is characterized by low-energy excitations in the magnetically ordered state with a Zeeman gap that scales with magnetic field $B$ =2$J_{ex}S$, which is indeed relevant in the present case [1, 59-63]. Notably, $S$ > 3/2 kagome lattices with Heisenberg exchange and strong easy axis or single ion anisotropy show a magnetization plateau [59-63]. A broad maximum around 0.8 K in ZF specific heat and Fisher-specific heat derived from DC magnetic susceptibility well above the transition temperature is close to that expected as per theoretical predictions $T/J_{ex}$~0.6 K for S=1/2 kagome magnets [57, 58], which evinces short-range spin-correlations in this frustrated magnet. The broad maximum in specific heat is sensitive to magnetic field, and an anomaly develops with an applied magnetic field of 0.9 T, which suggests that competing magnetic states are at play which are sensitive to external magnetic fields in the investigated frustrated magnet. The zero field μSR relaxation rate is governed by the Orbach mechanism typical of rare-earth-based frustrated magnets with CEF splitting with a gap that is consistent with thermodynamic and INS results. $^{11}$B NMR experiments unveil the presence of fluctuating electronic moments without the signature of magnetic ordering or freezing down to 1.8 K. The μSR asymmetry neither shows oscillations nor the 1/3 tail down to 34 mK ruling out the presence of long-range magnetic ordering and spin-freezing, respectively, which is at variance with that observed in specific heat. The absence of oscillations in muon asymmetry does not necessarily precludes long range magnetic order in this kagome magnet. One possible scenario could be that the co-existence of slow magnetic fluctuations along with static magnetism, and muon spin relaxation might be dominated by the slowly fluctuating magnetic moments at low temperatures. Another possible scenario for the absence of an anomaly at 0.3 K in μSR results could be due to the fact that the homogeneous internal field at the two muon stopping sites cancelled out. However, this seems unlikely, as both magnetically inequivalent sites would show this rather unusual effect simultaneously. A cancellation of dipolar fields arising from the ordered magnetic moments has indeed been observed, e.g., for a highly symmetric muon site in the heavy-fermion compound UPd$_2$Al$_3$ [74]. However, even in this case, a weak anomaly is observed at the transition temperature due to imperfections in the magnetic structure. One plausible scenario for the absence of oscillations in zero field muon asymmetry and no anomaly in μSR spin relaxation rate around 300 mK could be the fact that relaxation is possibly governed by the dipolar relaxation mechanism. Additionally, the structural disorder in the host spin-lattice could amplify the distribution of internal fields with a zero average value at the muon site, leading to the absence of oscillations in muon asymmetry as observed in several rare-earth based frustrated magnets. [74-81]. A frustrated 4$f$ magnet YbBO$_3$ shows magnetic ordering in specific heat and neutron diffraction at ~0.4 K which is ascribed to the co-existence of long range and short-range ordering [80] while μSR detects no magnetic ordering down to 20 mK, possibly related to structural disorder [81]. In the case of Nd$_3$BWO$_9$, the exact origin of the absence of oscillations in muon asymmetry is currently unclear that invokes further studies on high quality single crystals of Nd$_3$BWO$_9$. An interesting query is whether such an intriguing low-temperature magnetic state could harbor exotic fractional excitations in this frustrated magnet is not clear at present, but it provides a new direction to carry out detailed investigations on the single crystals of NBWO in the future [59,60].

**V. SUMMARY**

The frustrated magnet Nd$_3$BWO$_9$ with Nd$^{3+}$ moments embodying a distorted kagome spin-lattice is characterized by the interplay between spin-orbit coupling and spin correlations that leads to exotic magnetism and spin dynamics. The ground state Kramers doublet is well separated from the excited states, implying that the ground state can be described by $J_{eff}$ =1/2 spins, potentially leading to strong quantum fluctuations. The isothermal magnetization at sub-Kelvin temperature reveals a field-induced fractionalized magnetization plateau, suggesting an up-and-down ordered configuration of spins in each triangle of the frustrated kagome lattice with respect to the applied magnetic field. The specific heat, however, reveals the presence of a phase transition at 0.3 K that is attributed to the presence of interplane exchange interactions. The magnetic specific heat shows a broad maximum around 0.8 K, indicating the persistence of short-range spin correlations well above the magnetic phase transition temperature. The μSR relaxation at high temperatures is governed by an Orbach mechanism that is observed in several frustrated magnets based on rare earth ions. In addition, the μSR and NMR experiments reveal the presence of fluctuating electronic moments. The absence of a 1/3 tail in the muon asymmetry down to 34 mK rules out spin-freezing in this frustrated magnet, which is consistent with thermodynamic results. The absence of long-range magnetic ordering down to 34 mK, as revealed by our zero-field μSR experiments, suggests the predominant role of slowly fluctuating magnetic moments along with static magnetism at low temperatures. Furthermore, weaker dipolar relaxation mechanism, and structural disorder might explain the absence of oscillations in the zero-field μSR asymmetry. A clear picture concerning this warrants a thorough μSR investigation on the high-quality single crystals of $Nd_3BWO_9$. The external magnetic field drives several field-induced phases, implying the presence of a complex magnetic ordering phenomenon below 0.3 K in this spin-orbit-driven frustrated magnet. Our results suggest that competing magnetic states are active, which are sensitive to external magnetic fields, and that the low temperature physics is governed by static and fluctuating magnetic moments, along with short range spin correlations in this kagome antiferromagnet. Further studies on the single crystals focusing on the effect of external perturbations such as chemical and hydrostatic pressure may provide vital clues concerning the exotic magnetism and spin dynamics in this class of frustrated magnet. The current family of spin-orbit-driven rare-earth distorted kagome series $RE_3BWO_9$ offers a viable ground for the experimental realization of exotic quantum states and the establishment of a realistic microscopic Hamiltonian in this class of quantum materials.

**ACKNOWLEDGEMNTS**

P. K. acknowledges funding by the Science and Engineering Research Board and Department of Science and Technology, India through research grants. A. Y. thanks M. Barik for the help in synthesizing a batch of polycrystalline sample. A. Z. acknowledges the financial support of the Slovenian Research and Innovation Agency through the Programme No. P1-0125 and Projects Nos. J1-50008 and N1-0148. R. K. and A. E. acknowledge the financial support by DFG under Germany's Excellence Strategy EXC2181/1-390900948 (the Heidelberg STRUCTURES Excellence Cluster) and through the DAAD GSSP program D. T. A. acknowledges funding from EPSRC-UK (Grant No. EP/W00562X/1). We acknowledge PSI for beam time proposals. Experiments at the ISIS Neutron and Muon Source were supported by a beamtime allocation RB2220746 from the Science and Technology Facilities Council. Data is available here: https://doi.org/10.5286/ISIS.E.RB2220746 . We acknowledge the support of the HLD at HZDR, a member of the European Magnetic Field Laboratory.

# Appendix

### Inelastic Neutron Scattering: CEF parameters

In this appendix, we discuss the CEF parameters obtained from simultaneously fitting the INS spectra at 5 K and single crystal susceptibility data. The crystal structure of $Nb_3BWO_9$ is hexagonal, however, the point symmetry of the Nd ion is triclinic $C_1$. Due to this low point symmetry of Nd ion the CEF Hamiltonian has total of 27 independent CEF parameters. This makes it challenging to estimate all CEF parameters by fitting the INS data only. However, as we observed all 4 possible excitation peaks, we can use an variational method to fit the energy levels by varying the eigenvectors of the crystal field Hamiltonian [82]. Using a Monte Carlo procedure we obtained 16 sets of CEF parameters which fits the INS data well. The calculated magnetic susceptibility for all these parameter sets were almost identical and match the measured data well, but the calculated magnetization differed with different parameters yielding different saturation magnetization. We thus chose the parameter set which gave the saturation closest to that measured. Values of the obtained CEF parameters are given in the table below.

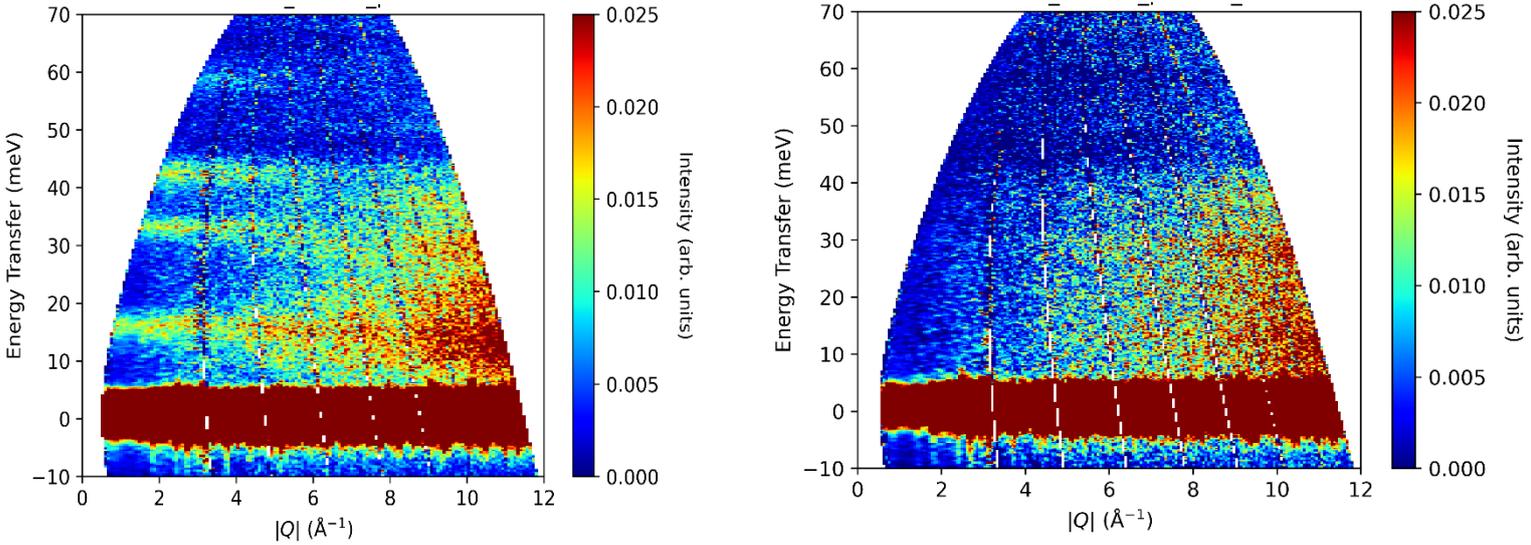

**Fig. S1** (a) Inelastic neutron scattering intensity of (a) $Nd_3BWO_9$ and (b) the non-magnetic reference $La_3BWO_9$. at 5 K for the incident energy $E_i$=80 meV. The four CEF excitation are clearly visible after subtraction of phonon contribution using the non-magnetic reference $La_3BWO_9$ as shown in Fig. 2 (a) of the main text.

### Table-II

The values of the CEF parameters in meV. The I before the parameters (i.e. $IB_2^1$) indicates that they are imaginary terms (sometimes called "sine" terms).

| $B_2^0$ = 0.0633 ( 0.037 ) | $B_6^1$ = 0.00087 (0.00012) | $IB_4^2$ = 0.01529 (0.0037) |
|---|---|---|
| $B_2^1$ = 0.2997 (0.032 ) | $B_6^2$ = 0.00100 (0.00019) | $IB_4^3$ = 0.05125 (0.012) |
| $B_2^2$ = -0.1193 ( 0.132 ) | $B_6^3$ = 0.00145 (0.00043) | $IB_4^4$ = -0.01905 (0.0029) |
| $B_4^0$ = -0.00105 (0.00037) | $B_6^4$ = 0.00055 (0.00013) | $IB_6^1$ = 0.00099 (0.00030) |
| $B_4^1$ = -0.0246 (0.002) | $B_6^5$ = -0.00468 (0.0011) | $IB_6^2$ = 0.00095 (0.00022) |
| $B_4^2$ = -0.00542 (0.0006) | $B_6^6$ = 0.00146 (0.00018) | $IB_6^3$ = 0.00202 (0.00019) |
| $B_4^3$ = 0.05049 (0.019) | $IB_2^1$ = -0.2500 (0.045) | $IB_6^4$ = -0.00067 (0.00026) |
| $B_4^4$ = -0.00462 (0.005) | $IB_2^2$ = -0.2409 (0.059) | $IB_6^5$ = 0.00024 (0.00029) |
| $B_6^0$ = -8.74e-5 (2.8e-5) | $IB_4^1$ = -0.01567 (0.0049) | $IB_6^6$ = -0.00151 (0.00039) |

## Table III

CEF energy levels (in meV) for the CEF model.

| $E_0$ | $E_1$ | $E_2$ | $E_3$ | $E_4$ |
|---|---|---|---|---|
| 0 | 15.983 | 33.085 | 42.696 | 58.738 |

## Table IV

The orthogonalized eigenstates $\pm\omega_n$ ($n$ =0-4) of the CEF Hamiltonian in the $|\pm m_J\rangle$ basis and the corresponding eigen energies of the $^4I_{9/2}$ Nd$^{3+}$ multiplet in the kagome magnet Nd$_3$BWO$_9$. Note that the coefficients of the second (denoted "-") level is the complex conjugate of the first (denoted "+").

| $\|\pm m_J\rangle$ | $\pm\omega_0$ | $\pm\omega_1$ | $\pm\omega_2$ | $\pm\omega_3$ | $\pm\omega_4$ |
|---|---|---|---|---|---|
| $\|\pm 9/2\rangle$ | -0.2108 + 0.1388i | 0.3361 - 0.2143i | -0.0145 - 0.3490i | -0.1366 + 0.2284i | 0.1816 - 0.2528i |
| $\|\pm 7/2\rangle$ | 0.0789 - 0.0444i | 0.2387 + 0.0589i | 0.1825 + 0.4934i | 0.1078 - 0.2271i | 0.0894 - 0.4466i |
| $\|\pm 5/2\rangle$ | -0.2452 + 0.0522i | -0.4643 + 0.3539i | 0.1669 - 0.2219i | -0.0631 + 0.0567i | -0.1009 - 0.1217i |
| $\|\pm 3/2\rangle$ | 0.0109 + 0.2281i | -0.2041 - 0.1794i | 0.3587 + 0.1479i | 0.1155 + 0.1594i | -0.1031 + 0.2261i |
| $\|\pm 1/2\rangle$ | -0.0920 + 0.0463i | -0.1276 + 0.2147i | -0.2903 + 0.2352i | -0.4476 + 0.2294i | 0.1961 - 0.0773i |
| $\|\mp 1/2\rangle$ | 0.1826 - 0.4044i | -0.1768 - 0.0653i | -0.0433 + 0.0793i | -0.4566 + 0.2022i | 0.0039 - 0.0052i |
| $\|\mp 3/2\rangle$ | 0.1564 + 0.5485i | -0.0926 - 0.0058i | 0.1719 - 0.2319i | -0.1273 + 0.0810i | 0.2487 - 0.3479i |
| $\|\mp 5/2\rangle$ | -0.2773 - 0.0335i | -0.1736 - 0.2561i | -0.2242 - 0.0037i | 0.1174 - 0.2930i | 0.2474 - 0.3197i |

| |∓7/2⟩ | 0.1528 + 0.2194i | 0.2013 + 0.1255i | 0.1193 + 0.1531i | -0.2958 - 0.2415i | 0.0957 + 0.2521i |
|---|---|---|---|---|---|
| |∓9/2⟩ | -0.2847 - 0.2228i | 0.2764 + 0.1763i | -0.0153 - 0.2333i | -0.1114 - 0.1926i | -0.2633 - 0.2758i |
| E (meV) | 0 | 15.983(1) | 33.085(1.5) | 42.696(2.1) | 58.738(4.04) |

_________________________________